\documentclass[11pt]{article}
\usepackage{moriond,epsfig}

\def\be{\begin{equation}}
\def\ee{\end{equation}}
\def\bea{\begin{eqnarray}}
\def\eea{\end{eqnarray}}

\newcommand{\newc}{\newcommand*}

\long\def\begincomment#1\endcomment{%
        \begingroup\sf\baselineskip12pt#1\endgroup}

\newc{\etal}{\textrm{et al.}} 
\newc{\eg}{\textrm{e.g.}} 
\newc{\ie}{\textrm{i.e.}}
\newc{\etc}{\textrm{etc}}
\newc\vs{\textrm{vs.}}
\newc{\cl}{\rm {CL}}

\newc{\ev}{\ensuremath{\,\mathrm{eV}}}
\newc{\kev}{\ensuremath{\,\mathrm{keV}}}
\newc{\mev}{\ensuremath{\,\mathrm{MeV}}}
\newc{\gev}{\ensuremath{\,\mathrm{GeV}}}
\newc{\tev}{\ensuremath{\,\mathrm{TeV}}}

\newc{\MeV}{\mev} 
\newc{\TeV}{\tev}
\newc{\invpb}{\ensuremath{/\text{pb}}}

\newc\nb{\ensuremath{\,\mathrm{nb}}} \newc\pb{\ensuremath{\,\mathrm{pb}}} \newc\fb{\ensuremath{\,\mathrm{fb}}}

\newc\pc{\ensuremath{\,\mathrm{pc}}}
\newc\kpc{\ensuremath{\,\mathrm{kpc}}}
\newc\mpc{\ensuremath{\,\mathrm{Mpc}}}

\newc\ps{\ensuremath{\,\mathrm{ps}}} 

\newc\cmeter{\ensuremath{\,\mathrm{cm}}} 
\newc\meter{\ensuremath{\,\mathrm{m}}} 
\newc\kmeter{\ensuremath{\,\mathrm{km}}}

\newc\second{\ensuremath{\,\mathrm{s}}}
\newc\msecond{\ensuremath{\,\mathrm{ms}}}
\newc\nsecond{\ensuremath{\,\mathrm{ns}}}
\newc\psecond{\ensuremath{\,\mathrm{ps}}}

\newc{\chisq}{\ensuremath{\chi^2}}
\newc{\chisqmin}{\ensuremath{\chi^2_{\mathrm{min}}}}
\newc{\Delchisq}{\ensuremath{\Delta\chi^2}}

\newc{\like}{\ensuremath{\mathcal{L}}}

\newc\lsim{\ensuremath{\mathrel{\rlap{\lower4pt\hbox{\hskip1pt$\sim$}}\raise1pt\hbox{$<$}}}}
\newc\gsim{\ensuremath{\mathrel{\rlap{\lower4pt\hbox{\hskip1pt$\sim$}}\raise1pt\hbox{$>$}}}}

\newc{\VEV}[1]{\ensuremath{\langle #1 \rangle}}

\newc{\dl}{\ensuremath{\stackrel{\leftarrow}{D}}}
\newc{\dr}{\ensuremath{\stackrel{\rightarrow}{D}}}

\newc{\bcenter}{\begin{center}}    \newc{\ecenter}{\end{center}}
\newc{\bfl}{\begin{flushleft}}    \newc{\efl}{\end{flushleft}}
\newc{\bfr}{\begin{flushright}}    \newc{\efr}{\end{flushright}}

\newc{\bi}{\begin{itemize}}
\newc{\ei}{\end{itemize}}
\newc{\bed}{\begin{description}}
\newc{\eed}{\end{description}}
\newc{\ben}{\begin{enumerate}}
\newc{\een}{\end{enumerate}}

\newc{\alphas}{\ensuremath{\alpha_s}}
\newc{\alphatwo}{\ensuremath{\alpha_2}}
\newc{\alphaone}{\ensuremath{\alpha_1}}
\newc{\alphai}[1]{\ensuremath{\alpha_{#1}}}
\newc{\alphaem}{\ensuremath{\alpha_{\mathrm{em}}}}

\newc{\alphaeff}{\ensuremath{\alpha_{\mathrm{eff}}}}

\newc{\sineff}{\ensuremath{\sin \theta_{\mathrm{eff}}}}
\newc{\sinsqeff}{\ensuremath{\sin^2 \theta_{\mathrm{eff}}}}
\newc{\dalphahad}{\ensuremath{\Delta \alpha_{\mathrm{had}}}}

\newc{\yt}{\ensuremath{h_t}} \newc{\yb}{\ensuremath{h_b}} \newc{\ytau}{\ensuremath{h_{\tau}}}

\newc\mz{\ensuremath{m_Z}} 
\newc\mw{\ensuremath{m_W}}
\newc\mZ{\mz}        \newc\mW{\mw}

\newc\mhsm{\ensuremath{ m_{H_{\mathrm{SM}}}}}

\newc{\mtop}{\ensuremath{ m_t}}               \newc{\mtpole}{\ensuremath{ M_t}}
\newc{\mbottom}{\ensuremath{ m_b}} 
\newc{\mtau}{\ensuremath{ m_{\tau}}}
\newc{\mt}{\mtpole}
\newc{\mb}{\mbottom} 

\newc{\llbar}{\ensuremath{\ell\bar{\ell}}}
\newc{\tauptaum}{\ensuremath{ \tau^+\tau^-}}

\newc{\qqbar}{\ensuremath{ q\bar{q}}} \newc{\ppbar}{\ensuremath{ p\bar{p}}}
\newc{\bbbar}{\ensuremath{ b\bar{b}}} \newc{\ttbar}{\ensuremath{ t\bar{t}}}
\newc{\ffbar}{\ensuremath{ f\bar{f}}} \newc{\tautaubar}{\ensuremath{ \tau\bar{\tau}}}

\newc{\mchi}{\ensuremath{m_\neutone}}
\newc{\squark}{\ensuremath{\tilde{q}}}
\newc{\slepton}{\ensuremath{\tilde{l}}}
\newc{\gluino}{\ensuremath{\tilde{g}}} 
\newc{\mgluino}{\ensuremath{{m_{\gluino}}}}

\newc{\sthw}{\ensuremath{ \sin\theta_W}}              \newc{\cthw}{\ensuremath{\cos\theta_W}}
\newc{\tanthw}{\ensuremath{ \tan\theta_W}}              \newc{\cotthw}{\ensuremath{\cot\theta_W}}

\newc{\ssqthw}{\ensuremath{\sin^2 \theta_W}}

\newc{\msbar}{\ensuremath{\overline{MS}}} \newc{\drbar}{\ensuremath{\overline{DR}}}

\newc{\mtmtsmmsbar}{\ensuremath{ m_t(m_t)^{\msbar}_{{\mathrm{SM}}}}}
\newc{\mtmtsmdrbar}{\ensuremath{ m_t(m_t)^{\drbar}_{{\mathrm{SM}}}}}
\newc{\mtmtmssmdrbar}{\ensuremath{ m_t(m_t)^{\drbar}_{{\mathrm{SUSY}}}}}

\newc{\mbmbmsbar}{\ensuremath{ m_b(m_b)^{\msbar} }}

\newc{\mbmbsmmsbar}{\ensuremath{ m_b(m_b)^{\msbar}_{{\mathrm{SM}}}}}
\newc{\mbmzsmmsbar}{\ensuremath{ m_b(\mz)^{\msbar}_{{\mathrm{SM}}}}}
\newc{\mbmzsmdrbar}{\ensuremath{ m_b(\mz)^{\drbar}_{{\mathrm{SM}}}}}
\newc{\mbmzmssmdrbar}{\ensuremath{ m_b(\mz)^{\drbar}_{{\mathrm{SUSY}}}}}

\newc{\mtaumzsmmsbar}{\ensuremath{ m_{\tau}(\mz)^{\msbar}_{{\mathrm{SM}}}}}
\newc{\mtaumzsmdrbar}{\ensuremath{ m_{\tau}(\mz)^{\drbar}_{{\mathrm{SM}}}}}
\newc{\mtaumzmssmdrbar}{\ensuremath{ m_{\tau}(\mz)^{\drbar}_{{\mathrm{SUSY}}}}}

\newc{\alphasmzms}{\ensuremath{\alpha_s(M_Z)^{\overline{MS}}}}
\newc{\alphaimzms}[1]{\ensuremath{\alpha_{#1}(M_Z)^{\overline{MS}}}}

\newc{\alphaemmz}{\ensuremath{\alpha_{\mathrm{em}}(M_Z)^{\overline{MS}}}}

\newc{\mzero}{\ensuremath{{m_0}}}
\newc{\mhalf}{\ensuremath{ m_{1/2}}}
\newc{\tanb}{\ensuremath{\tan\beta}}
\newc{\azero}{\ensuremath{ A_0}}
\newc{\signmu}{\ensuremath{\rm{sgn}\,\mu}}

\newc{\mgut}{\ensuremath{ M_{\rm GUT}}}
\newc{\mplanck}{\ensuremath{ M_{\rm P}}}      \newc{\mpl}{\ensuremath{ M_{\rm Pl}}}
\newc{\msusy}{\ensuremath{ M_{\rm SUSY}}}      \newc{\ms}{\ensuremath{ M_{\rm S}}}

 \newc{\hu}{\ensuremath{ H_u}}       \newc{\hd}{\ensuremath{ H_d}}
 \newc{\mhu}{\ensuremath{ m_{H_u}}}       \newc{\mhd}{\ensuremath{ m_{H_d}}}
 \newc{\mhuew}{\ensuremath{ m^{\ast}_{H_u}}}       \newc{\mhdew}{\ensuremath{ m^{\ast}_{H_d}}}
 \newc{\mhuewsq}{\ensuremath{ m^{\ast\, 2}_{H_u}}}       \newc{\mhdewsq}{\ensuremath{ m^{\ast\, 2}_{H_d}}}
 \newc{\mhl}{\ensuremath{m_\hl}} 
 \newc{\mglu}{\ensuremath{m_{\tilde g}}} 
 \newc{\mul}{\ensuremath{m_{\tilde{u}_L}}} 
 \newc{\mtone}{\ensuremath{m_{\tilde{t}_1}}} 

\newc{\sigsip}{\ensuremath{\sigma^{\rm SI}_{p}}}	\newc{\sigsin}{\ensuremath{\sigma^{\rm SI}_{n}}}
\newc{\sigsdp}{\ensuremath{\sigma^{\rm SD}_{p}}}	\newc{\sigsdn}{\ensuremath{\sigma^{\rm SD}_{n}}}
\newc{\sigsi}{\ensuremath{\sigma^{\rm SI}}}	\newc{\sigsd}{\ensuremath{\sigma^{\rm SD}}}

\newc{\abund}{\ensuremath{ \Omega h^2}}
\newc{\omegadm}{\ensuremath{ \Omega_{{\rm DM}}}}     \newc{\abunddm}{\ensuremath{ \Omega_{{\rm DM}} h^2}} 
\newc{\omegam}{\ensuremath{ \Omega_{{\rm m}}}}       \newc{\abundm}{\ensuremath{ \Omega_{{\rm m}} h^2}}
\newc{\omegab}{\ensuremath{ \Omega_{{\rm b}}}}	\newc{\abundb}{\ensuremath{ \Omega_{{\rm b}} h^2}}
\newc{\omegatot}{\ensuremath{ \Omega_{{\rm TOT}}}}
\newc{\omegacdm}{\ensuremath{ \Omega_{{\rm CDM}}}}   \newc{\abundcdm}{\ensuremath{ \Omega_{{\rm CDM}} h^2}}
\newc{\omegalambda}{\ensuremath{ \Omega_{\Lambda}}} \newc{\abundlambda}{\ensuremath{ \Omega_{\Lambda} h^2}}
\newc{\omegarad}{\ensuremath{ \Omega_{{\rm rad}}}}  \newc{\abundrad}{\ensuremath{ \Omega_{{\rm rad}} h^2}}

\newc{\rhocrit}{\ensuremath{ \rho_{\rm crit}}}
\newc{\rhochi}{\ensuremath{ \rho_{\chi}}}

\newc{\abunchi}{\ensuremath{\Omega_\chi h^2}}
\newc{\abundlsp}{\ensuremath{\Omega_{\rm LSP}h^2}}

\newc{\amu}{\ensuremath{ a_{\mu}}}        \newc{\amususy}{\ensuremath{ a_{\mu}^{\mathrm{SUSY}}}}
\newc{\amuexpt}{\ensuremath{ a_{\mu}^{\mathrm{expt}}}}        \newc{\amusm}{\ensuremath{ a_{\mu}^{\mathrm{SM}}}}
\newc\deltaamu{\ensuremath{\Delta a_{\mu}}} \newc{\deltaamususy}{\ensuremath{\delta a_{\mu}^{\mathrm{SUSY}}}}
\newc\gmtwo{\ensuremath{ (g-2)_{\mu}}} 
\newc{\deltagmtwomususy}{\ensuremath{\delta\left(g-2\right)_{\mu}^{\mathrm{SUSY}}}}
\newc{\deltagmtwomu}{\ensuremath{\delta\left(g-2\right)_{\mu}}}

\newc\BR{\ensuremath{\rm BR}}

\newc\bsgamma{\ensuremath{ b\rightarrow s \gamma }}
\newc\bxsgamma{\ensuremath{\overline{B}\rightarrow X_{s}\gamma}}

\newc\brbsgamma{\ensuremath{\BR\left(\bsgamma\right)}}
\newc\brbxsgamma{\ensuremath{\BR\left(\bxsgamma\right)}}

\newc\bsmumu{\ensuremath{B_s\to\mu^+\mu^-}}
\newc\brbsmumu{\ensuremath{\BR\left(B_s\to\mu^+\mu^-\right)}}

\newc\bdmmumu{\ensuremath{\overline{B}_d\to\mu^+\mu^-}}

\newc\bbbarmix{\ensuremath{\overline{B}_s\mbox{-}B_s}}      
\newc\delmbs{\ensuremath{\Delta M_{B_s}}}

\newc{\butaunu}{\ensuremath{\left(B_u \rightarrow \tau \nu\right)}}
\newc{\brbutaunu}{\ensuremath{\BR\left(B_u \rightarrow \tau \nu\right)}}

\newcommand*{\neutone}{\ensuremath{\chi}}

\let\oldcite\cite
\renewcommand*{\cite}{~\oldcite}

\newcommand*{\hl}{\ensuremath{h}}

\newcommand*{\ha}{\ensuremath{A}}

\begin{document}
\vspace*{4cm}
\title{Looking for supersymmetry: {\boldmath $\sim1\tev$} WIMP and the power\\ of complementarity in
  LHC and dark matter searches} 

\author{K. KOWALSKA, L. ROSZKOWSKI,~\footnote{Speaker at Rencontre de Moriond QCD and High Energy Phenomena session. On leave of absence from the University of Sheffield.} E.M. SESSOLO, S. TROJANOWSKI, A.J. WILLIAMS}

\address{BayesFITS Group, National Centre for Nuclear Research,\\ Ho{\. z}a 69, 
00-681 Warsaw, Poland}

\maketitle\abstracts{Some doubts have been expressed about low energy
  supersymmetry (SUSY) following the first run of the LHC. In this
  talk I will try to present a more upbeat view based on data, rather than theoretical expectations. In
  particular, I will make the following points: (a) in my opinion the
  most attractive candidate for dark matter (DM) is now the lightest
  neutralino with mass around 1\tev\ and with well defined properties
  (a nearly pure higgsino). This follows primarily from a combination
  of the properties of the discovered Higgs boson, in particular its
  mass close to 125\gev, and the relic density of dark matter, and is
  most clearly visible in the context of unified, or constrained, SUSY
  frameworks, although the DM candidate is quite generic; (b) this DM
  candidate will be nearly fully tested in forthcoming one-tonne DM
  underground search detectors; (c) the CMSSM will be nearly fully tested in the next few years by a
  combination of expected data from LHC experiments and from direct DM searches, as well as potentially also by the 
Cherenkov
  Telescope Array; (d) if naturalness plays any real role in SUSY
  (which remains unclear), then the amount of fine tuning, which
  is very large in simple models like the CMSSM, can be significantly
  reduced (even down to 1 in 20) with properly selected boundary
  conditions at the unification scale; (e) the \gmtwo\ anomaly can be 
  accommodated not only in the context of the general MSSM but also of some unified
  SUSY models with some superpartners to be within reach of the LHC.
}


The outcome of the first run of the LHC has brought some sense of
confusion and disappointment to
the ``new physics'' community. Not even a remotely convincing
hint of any signature of physics beyond the Standard Model (SM) emerged. In
particular, direct limits on superpartner masses were pushed up
significantly and, in the case of colored particles, reached and
exceeded the ballpark of $1\tev$, except for the stops. Likewise, in flavor violating processes measured at
the LHC and elsewhere SM predictions agree well with experimental
data, again implying that any potential new physics contributions
have to be suppressed, most likely by the large mass scale of new physics states. 
In particular, the measurements by LHCb and then also CMS of \brbsmumu,
both agreeing with  SM expectations suggest that the
pseudoscalar Higgs boson has to be heavy, at or beyond
$1\tev$, thus setting the scale for the heavy Higgs sector of the
MSSM. Note, however, that this does not apply to models with an extended Higgs sector, like the CNMSSM with an extra singlet Higgs, where some of the additional Higgs bosons can be much lighter, even lighter than the discovered Higgs boson.

These negative results (from the point of view of new physics searches) 
were actually independently reinforced by emerging properties of the
Higgs boson discovered by ATLAS and CMS. Its couplings to SM fermions and gauge bosons
came out to be SM-like, thus suggesting that the other MSSM Higgs bosons are
heavy, and decouple. 

Of course, one can take the ``I believe in what I can see'' approach
that the SM has been confirmed in light of the above and there is simply no new physics beyond the SM, at least up to the few TeV scale.  I believe that it is too quick to jump to such conclusions. Firstly, the
discovery of the Higgs boson is consistent not only with the SM but also with the
frameworks that predict a SM-like Higgs boson, in particular with the
MSSM.  Secondly, apart from its many well known theoretical puzzles (\eg, many
free parameters, apparently ad hoc gauge group or fermion
representations, \etc) the SM lacks explanations for big cosmological
questions: dark matter, baryogenesis, cosmic inflation, \etc.

Thirdly, within the framework of low energy SUSY, the mass of the Higgs
boson of around 125\gev\ is rather high and requires large radiative
corrections from the stop sector at, or above, the $1\tev$ scale. In SUSY in the context of grand unification (GUT), where soft SUSY breaking parameters are unified, this sets the scale for the SUSY
breaking scale \msusy\ in the multi-\tev\ regime, again consistent with direct SUSY search limits and flavor processes. In my opinion, this is a very important implication stemming from the properties of the discovered
Higgs boson that has perhaps not yet been appreciated widely enough
in the community.

\begin{figure}
\begin{center}
\psfig{figure=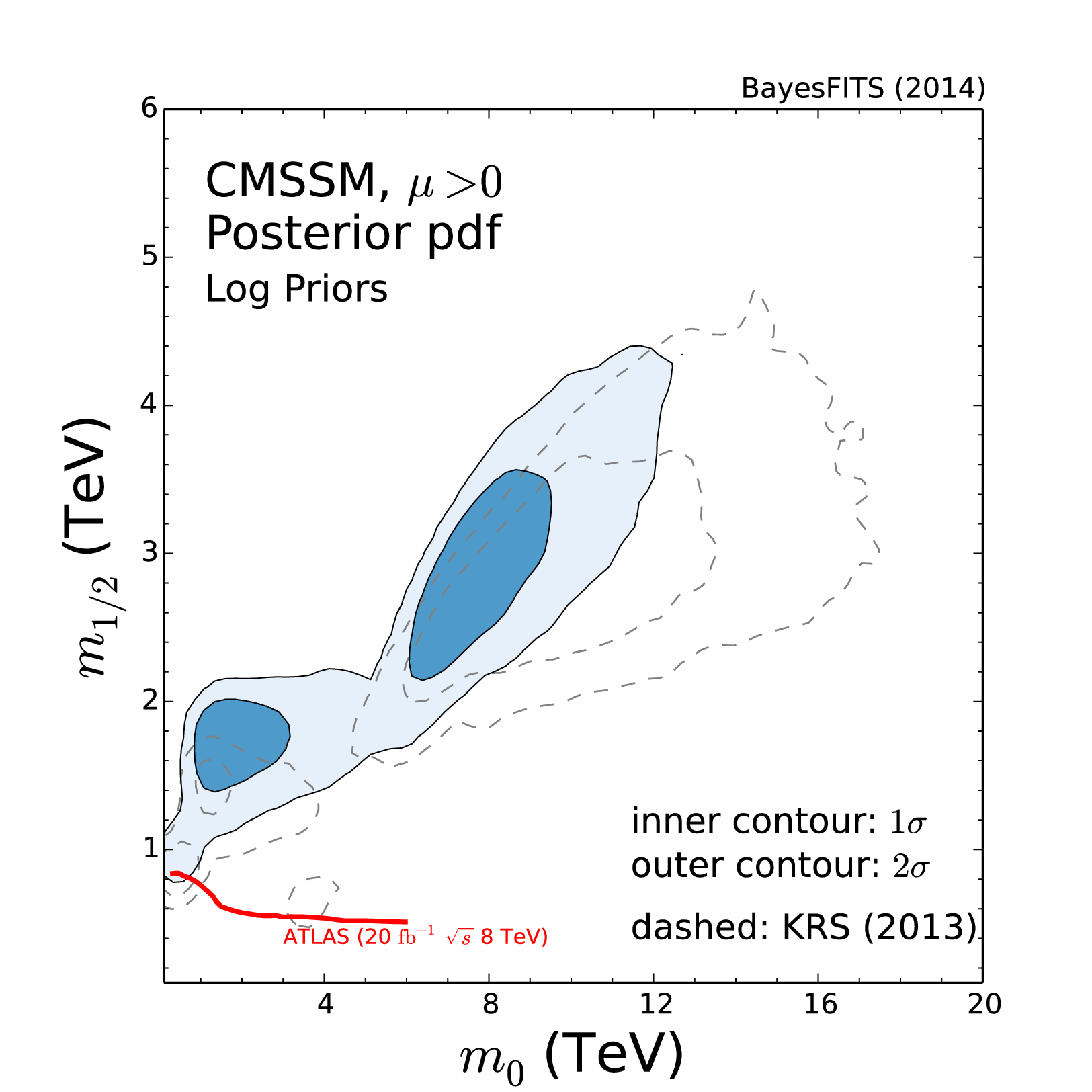,height=2.5in}
\hspace{3cm}
\psfig{figure=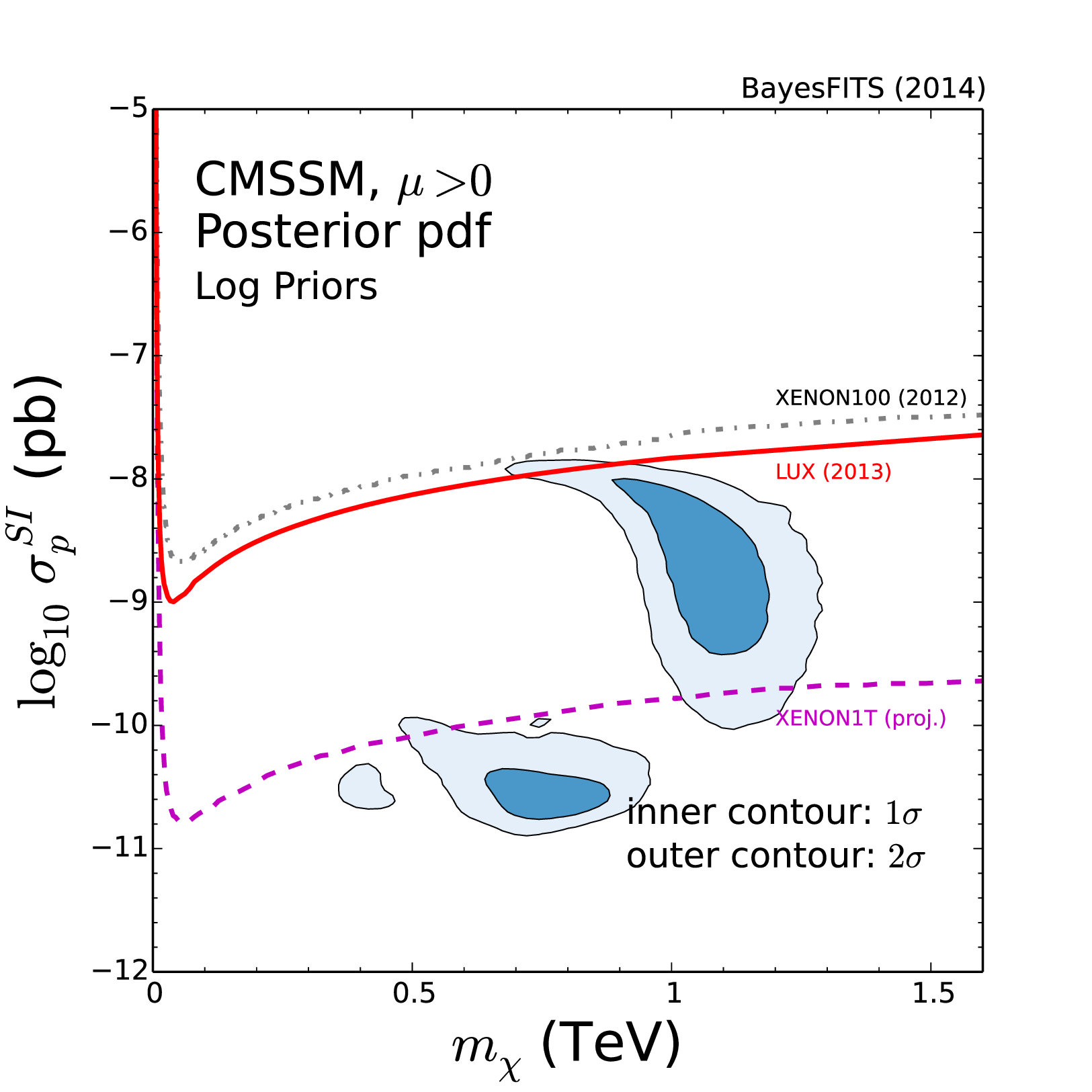,height=2.5in}
\caption{Left: Marginalized 2-dim. posterior pdf in the (\mzero,
  \mhalf) plane of the CMSSM for $\mu>0$, constrained by the
  experiments listed in Table~1 of Ref.\protect\cite{rsw14}.  The 68\% credible regions are
  shown in dark blue, and the 95\% credible regions in light blue. The
  solid red line shows the ATLAS combined 95\%~CL exclusion bound.
  Right: Marginalized 2D posterior pdf in the (\mchi,\sigsip) plane
  for the CMSSM constrained by the experiments listed in Table~1 of Ref.\protect\cite{rsw14}. The current LUX
  (solid red) and XENON100 (dash-dot gray) are also shown as well as   
  projected 1-tonne (dash magenta) limits.  Both figures are taken from Ref.\protect\cite{rsw14}.
  \label{fig:cmssmmupcase} }
\end{center}
\end{figure}

An additional important, and robust, constraint on SUSY parameter space is provided by
requiring that the relic density of the lightest neutralino assumed to
be DM agrees with the experimentally determined value in the
Universe. Since in constrained SUSY models this calculated quantity tends to be too large,
it can be satisfied only in some specific regions of parameter space. The outcome is
illustrated in the case of the CMSSM in the left panel of
Fig.~\ref{fig:cmssmmupcase} (taken from Ref.\cite{rsw14}, which is an
update of Ref.\cite{krs13}) where I present 1- and $2\sigma$ (credible) regions of
Bayesian total posterior probability (pdf).\footnote{Due to space limitation in this writeup I only summarize  some of our main results and cite only our relevant papers. I skip the description of the procedure and of the constraints adopted in deriving our numerical results. The reader is referred to our papers for a detailed presentation of our analysis and a list of references.} 

In the allowed bottom left corner, just above the red line marking the current
strongest limit from ATLAS, one can recognize a ``tiny''
stau-coannihilation region (SC), which appears at  only $2\sigma$,
followed, at larger \mzero\ and \mhalf, by an $\ha$-funnel (AF) region ($1\sigma$ region  occupying about $30\%$ of total Bayesian
probability). In both of these regions the neutralino is bino-like
which in the pre-LHC era was considered to be the most attractive WIMP
solution. 

Finally, at multi-TeV \mzero\ and \mhalf\ one can clearly see the
largest (about $70\%$) high probability region which I will call a
$\sim1\tev$ higgsino DM region (1TH). This is because in this region
the lightest neutralino is higgsino-like and its mass is set by the
$\mu$-parameter and is close to 1\tev\ in order to produce correct
relic abundance.

For comparison, in the left panel of
Fig.~\ref{fig:cmssmmupcase} dotted lines show the previously favored regions
based on the information available at that time (before 2013).\cite{krs13} In particular, Higgs mass calculation was performed at two 
loops (using FeynHiggs) and its experimental value taken in the analysis was higher, 125.8\gev\ (CMS). That is why the resulting ranges of 
\msusy\ were actually somewhat higher.

The $\sim1\tev$ higgsino DM region is a new region relative to the
pre-LHC studies of constrained SUSY models which explored much lower
values of \mzero\ and \mhalf, up to some 4\tev\ and 2\tev,
respectively,\cite{Fowlie:2012im} although the existence of such a region in constrained SUSY was
pointed out back in 2009, in the pre-LHC era, in the framework of the NUHM.\cite{Roszkowski:2009sm}  In fact, some initial reactions following the discovery of the Higgs
boson (whose mass at that time appeared even higher)
were to jump to conclusions that such a large value cannot be
accommodated in low energy SUSY. This is because in most
analyses \msusy\ was taken below a few \tev\ on the basis of the
theoretical expectation of ``naturalness''. I will comment on this
later.\footnote{It remains completely unclear to me whether the much emphasised criterion of naturalness in low-energy SUSY is actually not misguiding. I am more inclined to take a less theoretically motivated and more pragmatic view that can be put bluntly as: ``Natural is what is realized in Nature" - On a recent occasion I was very pleased to hear a very similar comment from Frank Wilczek.}

Remarkably, in this region the correct relic abundance
is achieved solely by the $\mu$-parameter being close to 1\tev. No special
mechanisms for reducing the relic density are required, unlike in the SC
or AF regions, nor does one invoke a contribution from several
unrelated (co)annihilation mechanisms. In this sense the 1TH region is
most natural. All one needs to do is to go to the regime of large enough
bino, and wino, mass. In
constrained models like the CMSSM this is achieved in regions of large
enough \mhalf.  It is worth mentioning that, by enlarging the ranges
of \mzero\ or \mhalf\ one does not find any new high probability regions. Also, in less constrained models like the NUHM the 1TH solution becomes even more pronounced than in the CMSSM.

In the likelihood function that we used in our Bayesian analysis
and numerical scans
all relevant constraints were actually included -- they are listed in
Table~1 of Ref.\cite{rsw14} where also the numerical procedure used in
performing our scans is described -- but it is primarily the Higgs
mass value and the DM relic abundance that determine the shape of the
favored regions. Some role is also played by direct limits on
superpartner masses (which we include in our likelihood function
through an approximate but accurate procedure described in Ref.\cite{krs13}) in setting a lower limit on \mhalf\ (and 
thus reducing the importance of the SC region to the mere $2\sigma$ level) and by
the updated measurements of \brbsmumu\ in giving more ``weight'' to the
AF region relative to a previous study\cite{krs13} and by current upper limits on \sigsip\ in basically ruling out the mixed neutralino DM region, characteristic of the hyperbolic branch/focus point region.

The result shown in Fig.~\ref{fig:cmssmmupcase} for the case of the
CMSSM is actually much more general. It applies both to unified, as well as
phenomenological models, with gaugino masses taken above the higgsino
mass of $\sim1\tev$.\cite{Roszkowski:2014iqa}

In the right panel of Fig.~\ref{fig:cmssmmupcase} the favored regions shown
in the left panel are mapped onto the (\mchi,\sigsip) plane. From
left to right, in the direction of increasing neutralino WIMP mass and
also \sigsip, we can see the ``tiny'' stau-coannihilation region (at
$2\sigma$ only), followed by the $\ha$-funnel region and, the
largest $\sim1\tev$ higgsino DM region. For comparison we show
also the current upper limits on \sigsip\ from LUX and Xenon100 which exclude the mixed neutralino DM region of HB/FP at a  few hundred GeV.

The emerging picture looks to me highly encouraging. The most attractive
$\sim1\tev$ higgsino DM region falls basically all within the reach of
upcoming one tonne detectors, like Xenon-1T which is expected to
produce new results by 2017 or so. This will provide the
most robust way of exploring this region.

Interestingly enough, this region should be independently probed
within expected sensitivity
reach of the forthcoming Cerenkov Telescope Array (CTA), a future
gamma-ray telescope due to start in 2018 or so, from observations of
diffuse radiation, assuming 500 hours of observation time plus, more importantly,
a steep enough DM density profile (close to the Einasto profile) towards the Galactic Center. (This being the case in simple constrained models, in phenomenological scenarios like the pMSSM there is, however, more freedom.\cite{Roszkowski:2014iqa})

\begin{figure}
\begin{center}
\psfig{figure=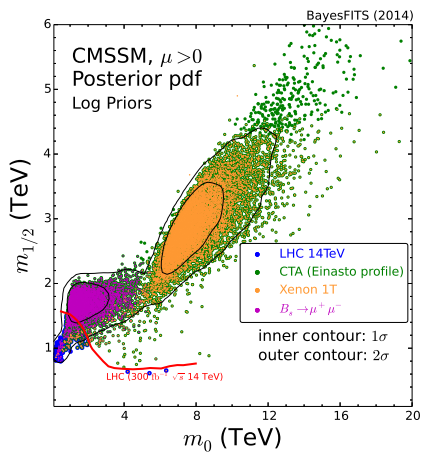,height=2.5in}
\vspace*{.4cm}
\caption{Complementarity of achivable experimental exploration of the
  CMSSM parameter space by LHC direct searches for SUSY (blue),
  \brbsmumu\ (magenta), one tonne DM search detectors (orange) and CTA
  (green). The solid inner and outher contours show the respective $68\%$ and $95\%$
  credible regions of the marginalized 2D posterior in the \mzero\ and
  \mhalf\ plane of the CMSSM with $\mu>0$. Taken from Ref.\protect\cite{rsw14}.
  \label{fig:cmssmcombo} }
\end{center}
\end{figure}

This can be seen in Fig.~\ref{fig:cmssmcombo} which also shows an
impressive complementarity of DM search experiments (both direct and
indirect in CTA) with LHC searches for signatures of SUSY. Both the SC
and the AF regions will be probably beyond the reach of one tonne DM
detectors but are expected to be accessible to the LHC14. The
first of them will hopefully be explored by direct detection searches
for SUSY. In the AF region, on the other hand, the superpartners are
too heavy to be detected at the LHC. Fortunately, a precise enough,
but achievable (at the level of $5-7\%$ of both experimental and
theory error) determination of \brbsmumu, if it comes out to be consistent with SM
predictions, would rule out most of the AF region. This is because in
this region $\tan\beta$ has to be large in order to enhance resonant
annihilation of DM by the $s$-channel heavy pseudoscalar Higgs
exchange but this, by the same token, also enhances the value of
\brbsmumu. Again, it is worth remembering, though, that in models with an extended Higgs sector, like the CNMSSM, the situation is more complicated and some of the Higgs bosons can be much lighter.

I will now briefly comment on two additional issues. The first is about so-called fine-tuning (FT) and is linked to ``naturalness". At $\msusy\gsim 1\tev$ FT is expected to be significant and indeed in simple constrained models like the CMSSM it is now very large, even in excess of 1 in 3000. Here I want to make a few points. Firstly, the validity of ``low fine-tuning" as a guiding principle in effective theories like low-energy SUSY remains (and always has been) unclear to me. We worry about the sensitivity of $\mz$ and other quantities at the EW scale to input parameters defined at the GUT scale only because we don't know the latter. In other words, we worry about FT because of our ignorance of physics at the high scale, while in fact FT in some effective theory may find its resolution in an underlying fundamental theory, or even in another effective, but more complete theory. A low energy analogue of this may be the GIM mechanism which provided a resolution to some divergencies in a three quark model by adding the charm quark to the picture. Secondly, following this spirt, by making a suitable choice of mass relations at the GUT scale among soft parameters and also by linking $\mu$ to the common scalar soft mass one can reduce FT in unified SUSY down to even 1 in 20.\cite{Kowalska:2014hza}

The final point is about the long-standing so-called \gmtwo\ anomaly which suggests that the measured value of \gmtwo\ is over $3\sigma$ above SM estimates. Explaining it in terms of SUSY would require low enough smuon and (at least one) neutralino and/or muon sneutrino and (at least one) chargino masses in order to provide large enough SUSY loop contributions to the quantity. It is therefore no surprise that one fails to reproduce  $\deltagmtwomu$ in simple constrained SUSY models where slepton masses are unified with those of squarks, while in the general MSSM (or its 19-dim subset called pMSSM) this can be easily done. Hence there is ``common wisdom" that constrained SUSY is incompatible with the \gmtwo\  anomaly. However, this is not quite true. One way is to simply disunify sleptons and squarks, another, less known, is to disunify gauginos. Some such possible constrained SUSY models with relaxed boundary conditions at the GUT scale have been shown to reproduce $\deltagmtwomu$ with the bonus that some light enough states must appear there and be for the most part accessible to LHC14. In other words, if the \gmtwo\ anomaly is real then either some light SUSY states will be seen at the LHC or those models will be ruled out.\cite{Kowalska:2015zja}

In conclusion, I've pointed out some distinct and well motivated phenomenological scenarios that can be put to a definitive experimental test at the LHC and in DM searches. While waiting for new data, and remaining open to surprises, I believe we have good reasons to remain optimistic. Long ago,  before the LHC era began, I formulated the conjecture: ``Low energy SUSY cannot be experimentally ruled out. It can only be discovered. Or else
abandoned.'' Indeed, while some specific SUSY models could in principle be excluded, I could not think of any experimental measurement that can be made in currently available facilities and that could rule out low energy SUSY as a framework. We should be able to know which way our field will go hopefully within the next few years.

\end{document}